\documentstyle[11pt, aaspp4, epsf, rotate]{article}

\begin{document}

\title{Early GRB Afterglows from Relativistic Blast Waves in General Radiative Regimes}

\author{Markus B\"ottcher\altaffilmark{1,2,3}
\and
Charles D. Dermer\altaffilmark{2}}

\altaffiltext{1}{Department of Space Physics and Astronomy, 
Rice University, MS 108, 6100 S. Main St., Houston, TX 77005-1892}

\altaffiltext{2}{E. O. Hulburt Center for Space Research,
Code 7653, Naval Research Laboratory,
Washington, D. C. 20375 - 5352}

\altaffiltext{3}{Chandra Fellow}

\centerline{\it Submitted to The Astrophysical Journal}

\begin{abstract} 
We present simple analytical expressions for the predicted
spectral and temporal behavior of the early afterglow radiation 
from gamma-ray bursts in radiative regimes intermediate
between the adiabatic and the fully radiative solutions of the
blastwave hydrodynamic equations. Our expressions are valid
as long as the relativistic electrons responsible for the
observed synchrotron emission are in the fast cooling regime 
and the blast wave is relativistic. We show that even a slight
deviation from a perfectly adiabatic evolution results in 
significant changes of the temporal characteristics of the 
afterglow emission.
\end{abstract}


\section{Introduction}

The relativistic blast wave model (\cite{mr92}, \cite{rm92}) 
has met with considerable success in explaining the X-ray, 
optical and radio afterglows of gamma-ray bursts (GRBs). 
In this model, the impulsive release of $\sim 10^{52}$ -- 
$10^{54}$ ergs of energy (e. g., \cite{kumar99}) into a 
small volume results in the formation of a relativistic 
blast wave. The initial Lorentz factor $\Gamma_0$ of 
material behind the shock wave reaches values of 
$\Gamma_0 \sim 100$ -- 1000 (\cite{bmk76}, \cite{cr78},
\cite{sp90}). The dominant radiation mechanism responsible for
the radio -- X-ray afterglows from these relativistic blast 
waves is believed to be optically thin synchrotron emission 
from relativistic electrons accelerated behind the shock 
front (\cite{mr93}, \cite{katz94}, \cite{tavani96}). 

With rapid follow-up observations of the X-ray, optical, and
radio afterglows of several GRBs (e. g., \cite{costa97}, \cite{piro98}, 
\cite{metzger97}, \cite{vanpar97}, \cite{djor97}, \cite{kul98},
\cite{frail97}) it has become possible to determine in great detail 
the spectral and temporal characteristics of these afterglows, 
which can then be compared to the predictions of the relativistic 
blast wave model (\cite{galama98b}, \cite{iwa99}).

The temporal evolution of flux levels and characteristic
frequencies of the synchrotron spectrum from electrons in
a relativistic blast wave can be formulated in a very elegant
way, directly comparable to observations, if the blast wave 
is either perfectly adiabatic or fully radiative (\cite{spn98}). 
However, while the comparison to afterglow data of several GRBs
suggests that at least in the late afterglow phase GRB
blast waves are well described by the adiabatic solution
(e. g., \cite{galama98b}), they can obviously not be 
perfectly adiabatic in order to be observable, and they
are generally believed to be strongly non-adiabatic in
the early afterglow phase. 

Analytic estimates similar to the ones found by Sari et al. 
(\markcite{spn98}1998) have been developed for intermediate 
radiative regimes under the a priori assumption that the 
blast wave evolution follows a self-similar behavior $\Gamma(r) 
\propto r^{-g}$ (\cite{rm98}, \cite{dcb99}, \cite{mr99}). 
However, the relation between the deceleration index $g$ 
and the blast wave energetics and radiative properties 
remains uncertain in these representations because the 
blastwave dynamical equations are not solved self-consistently.

Self-consistent solutions of the blast wave dynamics under
realistic assumptions on the energy transfer from protons
to electrons and on the magnetic field evolution, calculating
the synchrotron (and synchrotron self Compton) spectrum, solving
for the evolution of the electron distribution behind the shock
front and the blast wave dynamics simultaneously, have up to
now only been possible numerically (\cite{cd99}, \cite{huang99},
\cite{mod99}). 

In this paper, we show that under certain conditions, 
if the electrons are in the fast-cooling regime and the 
blast wave is relativistic, the blast wave kinetic equation 
can be solved analytically, yielding a simple, self-consistent 
analytical representation for the synchrotron emission from 
relativistic blast waves in general radiative regimes. In 
Section 2, we quote the blast wave kinetic equation and 
present its solution under the above assumptions. The 
spectral and temporal characteristics of the synchrotron
radiation from relativistic blast waves are derived in
Section 3. In Section 4, we discuss implications of a 
non-zero radiative efficiency for the interpretation of 
observed GRB afterglow characteristics. We summarize in 
Section 5. In this paper, we neglect the effects of 
synchrotron self-absorption and synchrotron-self Compton 
scattering. The importance of these effects is
investigated in detail in a separate paper (Dermer et
al., in preparation).

\section{The kinetic equation for general radiative regimes}

In the relativistic blast wave model it is assumed that a total
energy $E = 10^{52} \, E_{52}$~erg is deposited into a small
volume, giving rise to a relativistic blast wave expanding with
a bulk Lorentz factor $\Gamma_0$. The mass of the initial
baryon-loaded ejecta is given by $M_0 = E_0 / (\Gamma_0 \, 
c^2)$ in the comoving frame of the shocked material. As it
expands into an external medium of density $n_{ext} (r)
\propto r^{-\eta}$, it sweeps up matter at a rate $dm = 
\Omega \, r^2 \, n_{ext} (r) \, m_p \, dr$, where $\Omega$ 
is the solid angle element into which the blast wave is 
expanding. The kinetic equation governing the blast wave 
evolution is then given by (\cite{bmk76}, \cite{cd99})

\begin{equation}
{d\Gamma \over dm} = - {\Gamma^2 - 1 \over M},
\label{dGdm}
\end{equation}
where $M$ is the total, comoving mass (rest mass plus internal 
kinetic energy) of the ejecta + swept-up material. We assume
that a fraction $\epsilon_e$ of the swept-up energy per unit 
time is transferred to relativistic electrons behind the shock
front, and a fraction $\epsilon_{rad}$ of this energy is then
radiated as synchrotron radiation. Thus, a fraction $\epsilon
= \epsilon_e \epsilon_{rad}$ of the swept-up energy will be 
transformed into radiation. If the relativistic electrons are 
in the fast cooling regime (i. e. the synchrotron cooling time
scale for all electrons is shorter than the dynamical time scale), 
then the radiative efficiency $\epsilon_{rad} \approx 1$, and 
$\epsilon \approx \epsilon_e$ may be regarded as constant over 
the fraction of the blast wave evolution during which this 
condition is met. In this regime, Eq. \ref{dGdm} can be solved 
analytically. We note that up to now it is not certain whether
the condition $\epsilon_e \approx$~const. is actually met in
realistic blastwave scenarios. Investigation of this question
would require the detailed treatment of the process of energy
transfer from protons to electrons (see, e. g., Pohl \&
Schlickeiser \markcite{ps99}1999), which is beyond the scope
of this paper. Here, we assume that $\epsilon_e \approx$~const.
and note that our results may not be used if $\epsilon_e$
or $\epsilon_{rad}$ change significantly during the early 
afterglow phase.

As the blast wave propagates through the surrounding medium,
its mass (in the comoving frame) increases at a rate
\begin{equation}
dM = \left( \epsilon + \Gamma [ 1 - \epsilon] \right) \, dm.
\label{dM}
\end{equation}
Inserting Eq. \ref{dM} into Eq. \ref{dGdm}, we find 
\begin{equation}
M (\Gamma) = M_0 \left( {\Gamma^2 - 1 \over \Gamma_0^2 - 1} 
\right)^{-{(1 - \epsilon) \over 2}} \left( {\Gamma + 1 \over
\Gamma - 1} \right)^{\epsilon \over 2} \left( {\Gamma_0 + 1
\over \Gamma_0 - 1} \right)^{-{\epsilon \over 2}}.
\label{M}
\end{equation}
Using this relation in Eq. \ref{dGdm}, we have the general
solution
\begin{equation}
{m (r) \over M_0 \, A_0} = \int\limits_{\Gamma}^{\Gamma_0}
d\gamma \> {(\gamma + 1)^{\epsilon} \over (\gamma^2 - 1)^{3/2}},
\label{integral}
\end{equation}
where
\begin{equation}
m(r) \equiv \Omega \, m_p \, \int\limits_0^r d\tilde{r} \> 
\tilde{r}^2 \, n_{ext} (\tilde{r})
\label{mu}
\end{equation}
and
\begin{equation}
A_0 \equiv {\sqrt{\Gamma_0^2 - 1} \over \left( \Gamma_0 + 1 
\right)^{\epsilon}}.
\label{A_0}
\end{equation}
The integral on the right-hand-side of Eq. \ref{integral} can
be solved analytically in the extreme cases of an adiabatic
($\epsilon = 0$) and a fully radiative ($\epsilon = 1$) blast
wave, yielding
\begin{equation}
{m(r) \over M_0 \, A_0} = {\Gamma_{ad} \over 
\sqrt{\Gamma_{ad}^2 - 1}} - {\Gamma_0 \over \sqrt{\Gamma_0^2 - 1}}
\label{M_ad}
\end{equation}
and
\begin{equation}
{m(r) \over M_0 \, A_0} = {\Gamma_{rad} + 1 \over 
\sqrt{\Gamma_{rad}^2 - 1}} - {\Gamma_0 + 1\over \sqrt{\Gamma_0^2 - 1}}
\label{M_rad}
\end{equation}
in the adiabatic and radiative limits, respectively. It is 
straightforward to show that in the relativistic limit 
($\Gamma_0 > \Gamma \gg 1$) and in the deceleration phase 
($m(r) \gg M_0 / \Gamma_0$), these solutions approach the 
asymptotic behaviors $\Gamma_{ad} \propto r^{-(3 - \eta)/2}$ 
and $\Gamma_{rad} (r) \propto r^{- (3 - \eta)}$, if the 
external density is described by a radial profile
\begin{equation}
n_{ext} (r) = n_{r_0} \, \left( {r \over r_0} \right)^{-\eta}.
\label{n0}
\end{equation}
For general radiative regimes, the integral in Eq. \ref{integral}
can be solved in the relativistic limit, yielding
\begin{equation}
\Gamma_{rel} (r) = \Gamma_0 \, \left( 1 + [2 - \epsilon]
{m(r) \, \Gamma_0 \over \, M_0} \right)^{1 
\over \epsilon - 2},
\label{Gamma_rel}
\end{equation}
which has the asymptotic limits
\begin{equation}
\Gamma_{rel} (r) \approx \cases{ \Gamma_0 & if $(2 - \epsilon) \, 
m(r) \, \Gamma_0 \ll  M_0$ \cr \cr 
a \, r^{3 - \eta \over \epsilon - 2} & if $(2 - \epsilon) \, m(r) \,
\Gamma_0 \gg M_0$ \cr}
\label{rel_asymptote}
\end{equation}
where
\begin{equation}
a \equiv \left( {[3 - \eta] \, E_0 \, \Gamma_0^{-\epsilon} 
\over [2 - \epsilon] \, \Omega \, n_{r_0} \, m_p c^2 \, 
r_0^{\eta}} \right)^{1 \over 2 - \epsilon}.
\label{a}
\end{equation}
Being interested in the spectral characteristics of the afterglow
radiation, we will use the late-time asymptote of Eq. \ref{rel_asymptote}
in the remainder of this paper. Integrating over time $t$ in the
observer's frame, using $dt = dr/(\Gamma^2 c)$, we find
\begin{equation}
R (t) = \left( {a^2 \, c \, t \over \beta} \right)^{\beta}
\label{R}
\end{equation}
and
\begin{equation}
\Gamma (t) = a^{\beta} \, \left( {c \, t \over \beta} \right)^{\delta},
\label{Gamma}
\end{equation}
where
\begin{equation}
\beta = {2 - \epsilon \over 8 - 2 \eta - \epsilon}
\label{beta}
\end{equation}
and
\begin{equation}
\delta = {\eta - 3 \over 8 - 2 \eta - \epsilon} = {\beta - 1 \over 2}.
\label{delta}
\end{equation}
In the limits $\epsilon = 0$ and $\epsilon = 1$, Eqs. \ref{R} 
-- \ref{delta} reproduce the well-known scaling laws for
adiabatic and the radiative blast waves, respectively. 

\section{Temporal evolution of synchrotron spectra}

For the purpose of illustration, we will derive the relevant
scaling laws of the early afterglow emission here for the 
special case $\eta = 0$, i. e. for a homogeneous external 
medium of density $n_0$~cm$^{-3}$. A fraction $\epsilon_e$ 
($\approx \epsilon$ in the fast-cooling regime) of the 
swept-up energy is assumed to be transferred to a 
relativistic, nonthermal electron distribution injected into
the post-shock region with a power-law spectrum of index $p$
($n(\gamma) \propto \gamma^{-p}$), where we generally assume
$2 < p < 3$. With the energy normalization determined by the 
electron acceleration efficiency $\epsilon_e$, the electron 
injection distribution must have a low-energy cutoff at Lorentz 
factor (\cite{spn98})
\begin{equation}
\gamma_m \approx \epsilon_e \left( {p - 2 \over p - 1} \right)
\, {m_p \over m_e} \, \Gamma.
\label{gamma_m}
\end{equation}
The solution to the blast wave dynamics found in the previous 
Section is valid if the synchrotron loss time scale for electrons
at the low-energy cutoff $\gamma_m$ is shorter than the dynamical
time scale, which is comparable to the time elapsed after the
explosion. In this case, electrons injected prior to any given
time will have established a power-law spectrum with index $2$ with
a low-energy cutoff at
\begin{equation}
\gamma_c = {6 \, \pi \, m_e c \over \sigma_T \, \Gamma \, B^2 \, t}
\label{gamma_c}
\end{equation}
(\cite{spn98}), where the magnetic field $B$ is parametrized by a 
constant fraction $\epsilon_B^{1/2}$ of the equipartition value,
\begin{equation}
B = \sqrt{ 32 \, \pi \, m_p c^2 \, \epsilon_B \, n_{ext}} \, \Gamma,
\end{equation}
and $\gamma_c < \gamma_m$ for the electrons to be in the fast cooling
regime. It is conceivable that the magnetic-field equipartition parameter 
$\epsilon_B$ is actually not constant, but increases with time. This 
would not alter Eqs. \ref{F_c} -- \ref{fnuc} (although it would obviously
introduce an additional time-dependence through $\epsilon_B$), but would 
expand the period of validity of our results because the electrons would 
remain in the fast-cooling regime over a longer fraction of the early
blast wave evolution.

At any given time, the spectrum of electrons behind the shock
will then be given by a low-energy cutoff at $\gamma_c$, a power-law 
with index $2$ for $\gamma_c < \gamma < \gamma_m$ and a second power-law
with index $p + 1$ for $\gamma > \gamma_m$. The synchrotron spectrum
generated by such an electron distribution (neglecting synchrotron-self
absorption) consists of a double-broken power-law given by
\begin{equation}
F_{\nu} = F_c \, \cases{ (\nu / \nu_c)^{1/3} & for $\nu < \nu_c$, \cr
(\nu / \nu_c)^{-1/2} & for $\nu_c < \nu < \nu_m$, \cr
(\nu_m / \nu_c)^{-1/2} \, (\nu / \nu_m)^{-p/2} & for $\nu > \nu_m$ \cr}
\label{Fnu}
\end{equation}
where the characteristic synchrotron frequencies $\nu_i = (m_e c^2 / h) \,
(B / B_{cr}) \gamma_i^2$ with $B_{cr} = 4.414 \cdot 10^{13}$~G and $F_c$ 
is the flux at frequency $\nu_c$. 

Now, normalizing the total number of swept-up, relativistic electrons
in the shock to $N_e = (4/3) \pi n_{ext} R^3$, and the total radiative
power to $L = (4/3) \, \Gamma \, c \, \sigma_T \, (B^2 / [8 \pi]) 
\int_{\gamma_c}^{\infty} N_e (\gamma) \, \gamma^2 \, d\gamma$, we
find
\begin{equation}
F_c = {B \, B_{cr} \over 36 \, \pi \, d_L^2} \, {h \over m_e c} \, 
\sigma_T \, \Gamma \, R^3 \, n_{ext},
\label{F_c}
\end{equation}
where $d_L = 10^{28} \, d_{28}$~cm is the luminosity distance to the 
GRB source. 

We can now use Eqs. \ref{R} -- \ref{F_c} to find the relevant scaling
laws for the flux normalization $F_c$ and the break frequencies $\nu_c$
and $\nu_m$. Denoting by $t_d$ the time in the observer's frame in 
days, we find

\begin{equation}
F_c = f_{F_c} \> d_{28}^{-2} \, \epsilon_B^{1/2} \, \left( {E_{52} 
\over \Omega} \right)^{\phi_E} \, \Gamma_0^{-\epsilon \, \phi_E} \, 
n_0^{\phi_n} \> t_d^{\phi_t} \; {\rm Jy}
\label{Fc_scale}
\end{equation}
with
\begin{equation}
f_{F_c} = 2.51 \cdot 10^{(-56.0 + 15.4 \, \phi_t + 55.3 \, \phi_E)} \,
(8 - \epsilon)^{\phi_t} \, (2 - \epsilon)^{-(\phi_t + \phi_E)},
\label{fFc}
\end{equation}

\begin{equation}
\nu_m = {f_{\nu_m} \over (1 + z)} \epsilon_B^{1/2} \, \epsilon_e^2 
\, \left( {E_{52} \over \Omega} \right)^{\chi_E} \, \Gamma_0^{-\epsilon 
\, \chi_E} \, n_0^{\chi_n} \> t_d^{\chi_t} \; {\rm Hz}
\label{num_scale}
\end{equation}
with
\begin{equation}
f_{\nu_m} = 3.74 \cdot 10^{(12.0 + 15.4 \, \chi_t + 55.3 \, \chi_E)}
\, (8 - \epsilon)^{\chi_t} \, (2 - \epsilon)^{-(\chi_t + \chi_E)}
\, \left( {p - 2 \over p - 1} \right)^2,
\label{fnum}
\end{equation}
and

\begin{equation}
\nu_c = {f_{\nu_c} \over (1 + z)} \, \epsilon_B^{-3/2} \, \left( 
{E_{52} \over \Omega} \right)^{\psi_E} \, \Gamma_0^{-\epsilon \, 
\psi_E} \, n_0^{\psi_n} \> t_d^{\psi_t} \; {\rm Hz}
\label{nuc_scale}
\end{equation}
with
\begin{equation}
f_{\nu_c} = 2.40 \cdot 10^{(46.0 + 15.4 \, \psi_t + 55.3 \, \psi_E)}
\, (8 - \epsilon)^{2 + \psi_t} \, (2 - \epsilon)^{-(2 + \psi_t + \psi_E)}.
\label{fnuc}
\end{equation}
The power-indices of these relations are given in Table \ref{indices}.
At a given observing frequency $\nu_{obs}$ the flux will decay according
to power-laws in time, depending on the part of the synchrotron spectrum 
containing the observing frequency at a given time. For $\nu < \nu_c$
we have $F_{\nu} \propto \nu^{1/3} \, t^{\omega_0}$ with
\begin{equation}
\omega_0 = \phi_t - {\psi_t \over 3} 
= {1 \over 3} \, {4 - 11 \, \epsilon \over 8 - \epsilon}.
\label{omega0}
\end{equation}
For $\nu_c < \nu < \nu_m$ we have $F_{\nu} \propto \nu^{-1/2} \,
t^{\omega_1}$ with
\begin{equation}
\omega_1 = \phi_t + {\psi_t \over 2}
= - 2 \, {1 + \epsilon \over 8 - \epsilon}.
\label{omega1}
\end{equation}
For $\nu > \nu_m$, $F_{\nu} \propto \nu^{-p/2} \, t^{\omega_2}$ with
\begin{equation}
\omega_2 = \phi_t + {\psi_t \over 2} + {(p - 1) \over 2} \, \chi_t
= - {2 \, (1 + \epsilon) + 6 \, (p - 1) \over 8 - \epsilon}.
\label{omega2}
\end{equation}
Eqs. \ref{num_scale}, and \ref{nuc_scale} are easily inverted to find 
the sweep-through times $t_c$ and $t_m$ of the break frequencies $\nu_c$ 
and $\nu_m$ at a given observing frequency, which define the times of 
breaks in the light curves between the above power-law decay slopes. 
We have $t_c \propto \nu^{1/\psi_t}$ and $t_m \propto \nu^{1/\chi_t}$.

From the indices in Tab. \ref{indices} one sees that generally the
break frequency $\nu_m$ decreases more rapidly with time than the 
cooling frequency $\nu_c$. If synchrotron self Compton scattering
is unimportant (see Dermer et al., in preparation), this 
means that a burst which starts out in the fast-cooling regime 
will become radiatively less efficient, with a time-dependent 
$\epsilon$, after both break frequencies have become equal. The 
time $t_{d, cm}$ (in days) at which this happens, and which 
terminates the phase of applicability of our results, is given 
by

$$
t_{d, cm} = 10^{\left(- {33.8 \over \tau_{cm}} - 15.41 + 
{221 \over [4 + \epsilon]} \right)} (8 - \epsilon)^{- {12 \over 4 + 
\epsilon}} \, (2 - \epsilon)^{8 \over 4 + \epsilon}
$$
\begin{equation}
\cdot \> ( \epsilon_B \, \epsilon_e )^{2 \over \tau_{cm}} \, 
\left( {E_{52} \over \Omega} \right)^{4 \over 4 + \epsilon} \, 
\Gamma_0^{- {4 \epsilon \over 4 + \epsilon}} \, n_0^{4 - 
\epsilon \over 4 + \epsilon} \, \left( {p - 2 \over p - 1}
\right)^{2 \over \tau_{cm}},
\label{tcm}
\end{equation}
where $\tau_{cm} \equiv (8 + 2 \epsilon) / (8 - \epsilon)$.
After this transition, the radiative efficiency will steadily
decrease as $\epsilon \approx \epsilon_e \, (\gamma_m / 
\gamma_c)^{p - 2}$ (\cite{mod99}) until the blast wave 
evolution approaches the adiabatic limit ($\epsilon \to 
0$) in which, e. g. the expressions given by Sari et al. 
(\markcite{spn98}1998) may be applied as long as the 
blast wave is relativistic.

\section{The influence of $\epsilon > 0$}

Fig. \ref{lightcurves} shows the GRB afterglow light curves for
a standard GRB (see figure caption for parameters) at infrared,
optical, and soft X-ray frequencies during the relativistic and
fast-cooling phase of the blast wave (Eqs. \ref{Fc_scale} -- 
\ref{nuc_scale} are no longer applicable if the blast wave 
becomes non-relativistic or is in the slow-cooling regime). The 
figure illustrates that there are very pronounced differences between 
the light curves at a given frequency even for slight changes in the
radiative efficiency. In particular, the rise/decay slope $\omega_0$,
when the observing frequency $\nu$ is in the $F_{\nu} \propto \nu^{1/3}$ 
part of the spectrum, and the cooling-break sweep-through time $t_c$
are seen to depend very strongly on the radiative regime. 

The dependence of the temporal rise/decay slopes $\omega_i$ on the 
radiative regime, as given by eqs. \ref{omega0} -- \ref{omega2} is 
illustrated in Fig. \ref{omega} for our standard set of GRB parameters. 
Both the slopes $\omega_0$ and $\omega_2$ (corresponding to the 
$\nu^{1/3}$ and the $\nu^{-p/2}$ parts of the spectrum, respectively) 
are rapidly decreasing with increasing radiative efficiency $\epsilon$. 
We also see that the typically observed power-law decays of X-ray and 
optical afterglows, $F_{\nu} \propto t^{-\alpha}$ with $1.1 \lesssim 
\alpha \lesssim 1.4$ (e. g., \cite{costa97}, \cite{piro98}, \cite{zand98},
\cite{galama98a}, \cite{diercks98}), if produced by a blast wave 
in the fast-cooling regime, imply additional constraints on a rather 
hard electron injection spectrum if one allows for a finite radiative 
efficiency $\epsilon \gtrsim 0.1$. 

In Fig. \ref{breaktimes} we plot the sweep-through times of the
break frequencies $\nu_c$ and $\nu_m$ for observations at optical, 
infrared, and soft X-ray frequencies for our standard burst parameters 
as a function of the radiative efficiency. In particular for low 
$\epsilon$, both break times depend very strongly on the radiative 
efficiency. Thus any conclusions drawn from the observed sweep-through 
times based on the adiabatic solution of the blast wave dynamics, 
need to be taken with caution. Note that in calculating these break 
times as well as the light curves shown in Fig. \ref{lightcurves}, 
we have set $\epsilon_e = \epsilon$, which is the reason for the 
sweep-through time $t_m$ approaching $0$ for $\epsilon \to 0$. 
In the adiabatic limit $\epsilon \to 0$ this always leads to 
$t_m < t_c$, implying $\nu_m < \nu_c$, where our solution is 
not applicable. The figure also shows the transition time 
$t_{d, cm}$, at which $\nu_c = \nu_m$.

Fig. \ref{transition} illustrates an example of how the 
frequency-dependent GRB afterglow light curves are expected 
to change between the early afterglow phase investigated in 
this paper, and the later, most probably adiabatic phase. 
For the example shown in the figure, we have assumed that 
the blast wave starts out in the fast-cooling regime with 
$\epsilon = \epsilon_e = 0.8$. The other parameter values are
given in the figure caption. According to Eq. \ref{tcm}, 
our solutions are valid until $\sim 35$~h into the afterglow 
phase. The following transition phase, in which the blast 
wave still has a non-negligible, but steadily decreasing 
radiative efficiency, can up to now only be treated numerically. 
In the late afterglow phase, the light curves asymptotically 
approach the adiabatic evolution, for which we are showing 
the results of Sari et al. (\markcite{spn98}1998). 

\section{Summary and conclusions}

We have presented an analytic solution to the hydrodynamic 
evolution of a relativistic blast wave and the broad-band 
synchrotron spectrum of electrons accelerated behind the 
shock front in general radiative regimes under the assumption
that the electrons are in the fast cooling regime and the electron
acceleration efficiency $\epsilon_e$ and the magnetic-field
equipartition parameter $\epsilon_B$ are constant. We have
shown that in this regime small deviations from a perfectly
adiabatic evolution will have strong, potentially observable 
effects on the temporal evolution of the early afterglow 
radiation. In particular, the rise/decay slopes of the 
monochromatic flux and the times of expected breaks in 
the light curves due to spectral break frequencies sweeping 
past the observing frequency, are significantly affected by 
a non-zero radiative efficiency. 

Our predictions will be particularly important for the interpretation
of rapid optical -- X-ray follow-up observations of early ($\sim$~a 
few hours after the burst) GRB afterglows, which might become
possible with much improved data quality with the upcoming HETE~II 
satellite and with the planned SWIFT mission and with high-sensitivity,
high-spectral-resolution follow-up observations by the new generation
of X-ray telescopes such as Chandra or XMM. By comparison of the 
decay laws of the afterglow radiation in different spectral regimes 
and of the shift of the break frequencies with time with our eqs. 
\ref{omega0} -- \ref{omega2} and the indices given in Table
\ref{indices}, the radiative regime of the blast wave can be
determined unambiguously. Having fixed the radiative efficiency
$\epsilon$, eqs. \ref{Fc_scale}, \ref{num_scale}, and \ref{nuc_scale}
can be used to determine $\epsilon_B$, $\Gamma_0$, and $n_0$, if
one has an independent estimate of the total energy $E_{52}$.
If such an estimate is not available, one would need an additional
equation, which can be provided by the location of the synchrotron 
self absorption frequency, in order to solve the system of equations
for all four free parameters ($\epsilon_B$, $\Gamma_0$, $n_0$, and
$E_{52}$).

\acknowledgements{This work was supported by NASA through grand
NAG~5-4055 and the Chandra Postdoctoral Fellowship grant number
PF~9-10007, awarded by the Chandra X-ray Center, which is operated
by the Smithsonian Astrophysical Observatory for NASA under
contract NAS~8-39073.}

\newpage

\begin{table}
\caption[]{Indices for time, energy, and density scaling of the
flux $F_c$, and the break frequencies $\nu_m$ and $\nu_c$}
\label{indices}
\begin{center}
\begin{tabular}{ccc}
\hline
$F_c$             & $\nu_m$ &               $\nu_c$ \\
\hline
$\phi_t = - {3 \epsilon \over 8 - \epsilon}$ & $\chi_t = - {12 \over
8 - \epsilon}$ & $\psi_t = - 2 {2 - \epsilon \over 8 - \epsilon}$ \\
 & & \\
$\phi_E = {8 \over 8 - \epsilon}$ & $\chi_E = {4 \over 8 - \epsilon}$
& $\psi_E = - {4 \over 8 - \epsilon}$ \\
 & & \\
$\phi_n = {3 \over 2} - \phi_E$ & $\chi_n = {1 \over 2} - \chi_E$ &
$\psi_n = -{3 \over 2} - \psi_E$ \\
 & & \\
\hline
\end{tabular}
\end{center}
\end{table}

\newpage

\begin{figure}
\epsfysize=11cm
\rotate[r]{
\epsffile[0 40 600 470]{figure1.ps}}
\caption[]{GRB afterglow light curves in different radiative
regimes at infrared (upper panel), optical (middle panel),
and soft X-ray (lower panel) frequencies. Parameters:
$E_{52} / \Omega = 100/(4\pi)$, $n_0 = 100$, $\Gamma_0 = 300$, 
$\epsilon_B = 10^{-2}$, $\epsilon_e = \epsilon$, $p = 2.5$, $z = 1$. }
\label{lightcurves}
\end{figure}

\newpage

\begin{figure}
\epsfysize=11cm
\rotate[r]{
\epsffile[50 70 550 500]{figure2.ps}}
\caption[]{The temporal decay index $\omega_i$ in the three
different spectral phases as a function of the radiative 
efficiency $\epsilon$. The index $\omega_2$ (decay in the 
$\nu^{-p/2}$ part of the synchrotron spectrum) is shown for 
two different values of the electron injection spectral 
index $p$.}
\label{omega}
\end{figure}

\newpage

\begin{figure}
\epsfysize=11cm
\rotate[r]{
\epsffile[50 70 550 500]{figure3.ps}}
\caption[]{The sweep-through times of the cooling and the
$\nu_m$ break as a function of the radiative efficiency
$\epsilon$ for three different observing frequencies. Parameters:
$E_{52} / \Omega = 100 / (4\pi)$, $n_0 = 10$, $\Gamma_0 = 100$, 
$\epsilon_B = 0.2$, $\epsilon_e = \epsilon$, $p = 2.5$, $z = 1$. 
Also shown is the time $t_{cm}$, at which the transition from
fast cooling to slow cooling occurs ($\nu_c = \nu_c$). The 
cooling-break sweep-through times $t_c$ for $\nu_{obs} = 
3 \cdot 10^{14}$~Hz and $10^{17}$~Hz are of the order 
$\lesssim 1$~s and thus irrelevant since this is shorter 
than the deceleration time scale.}
\label{breaktimes}
\end{figure}

\newpage

\begin{figure}
\epsfysize=11cm
\rotate[r]{
\epsffile[0 40 600 470]{figure4.ps}}
\caption[]{Transition of the GRB afterglow light curves 
at infrared, optical, and soft X-ray frequencies from the
fast-cooling, quasi-radiative regime to the slow-cooling,
adiabatic regime. Parameters: $E_{52} / \Omega = 100/(4\pi)$, 
$n_0 = 100$, $\Gamma_0 = 300$, $\epsilon_B = 10^{-2}$, 
$\epsilon_e = \epsilon = 0.8$ for $t < t_{cm}$, $p = 2.5$, 
$z = 1$.}
\label{transition}
\end{figure}

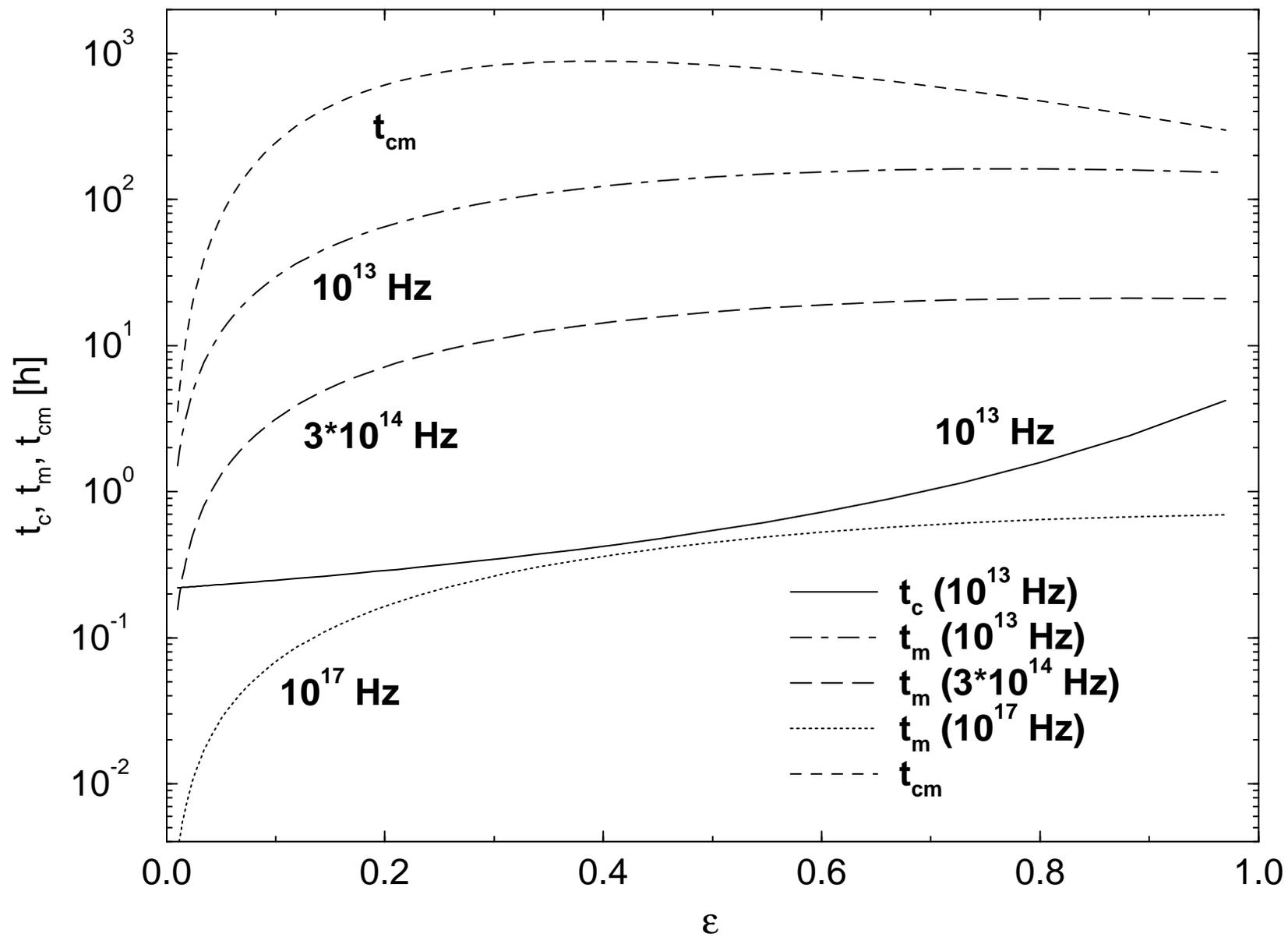
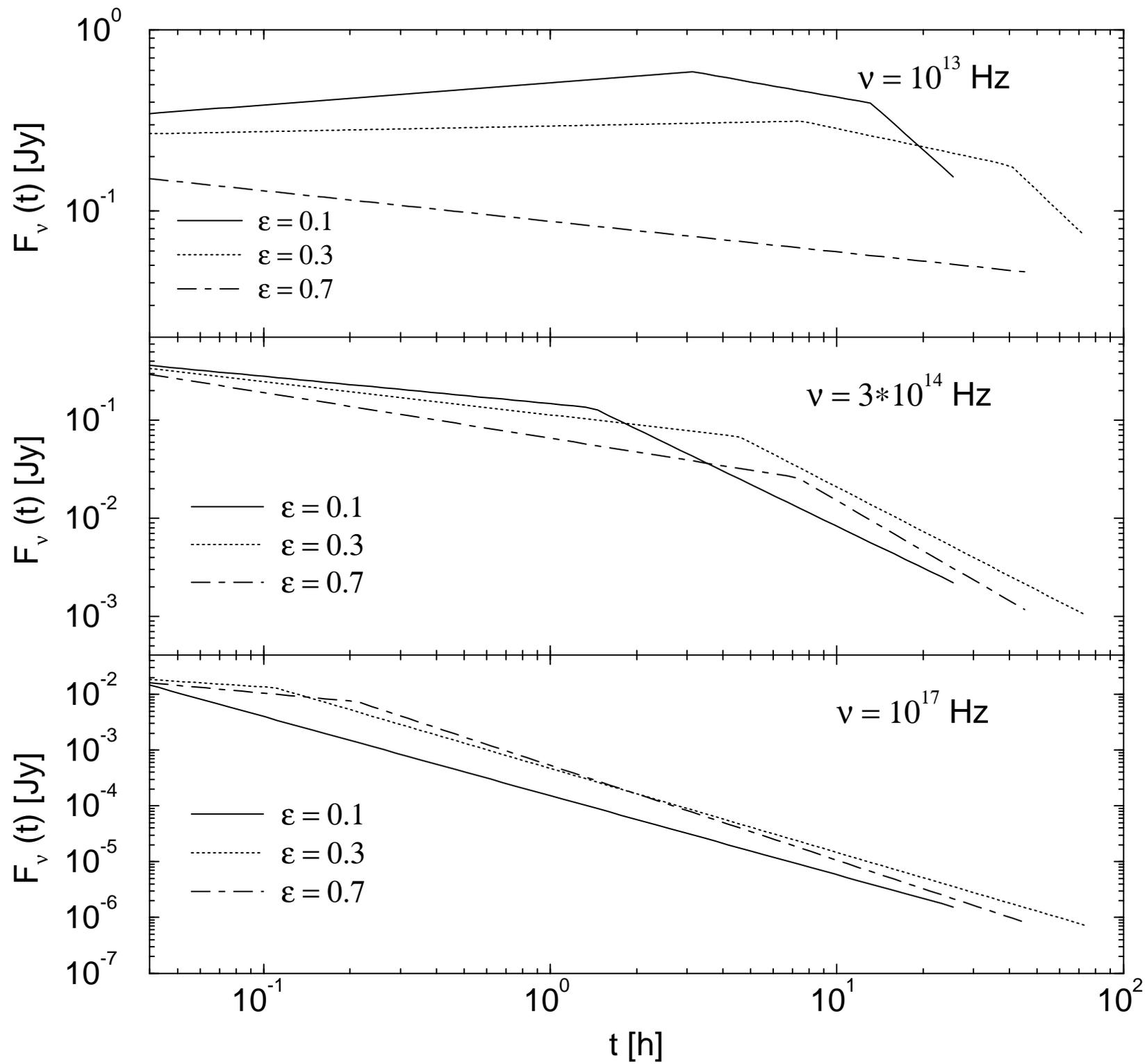
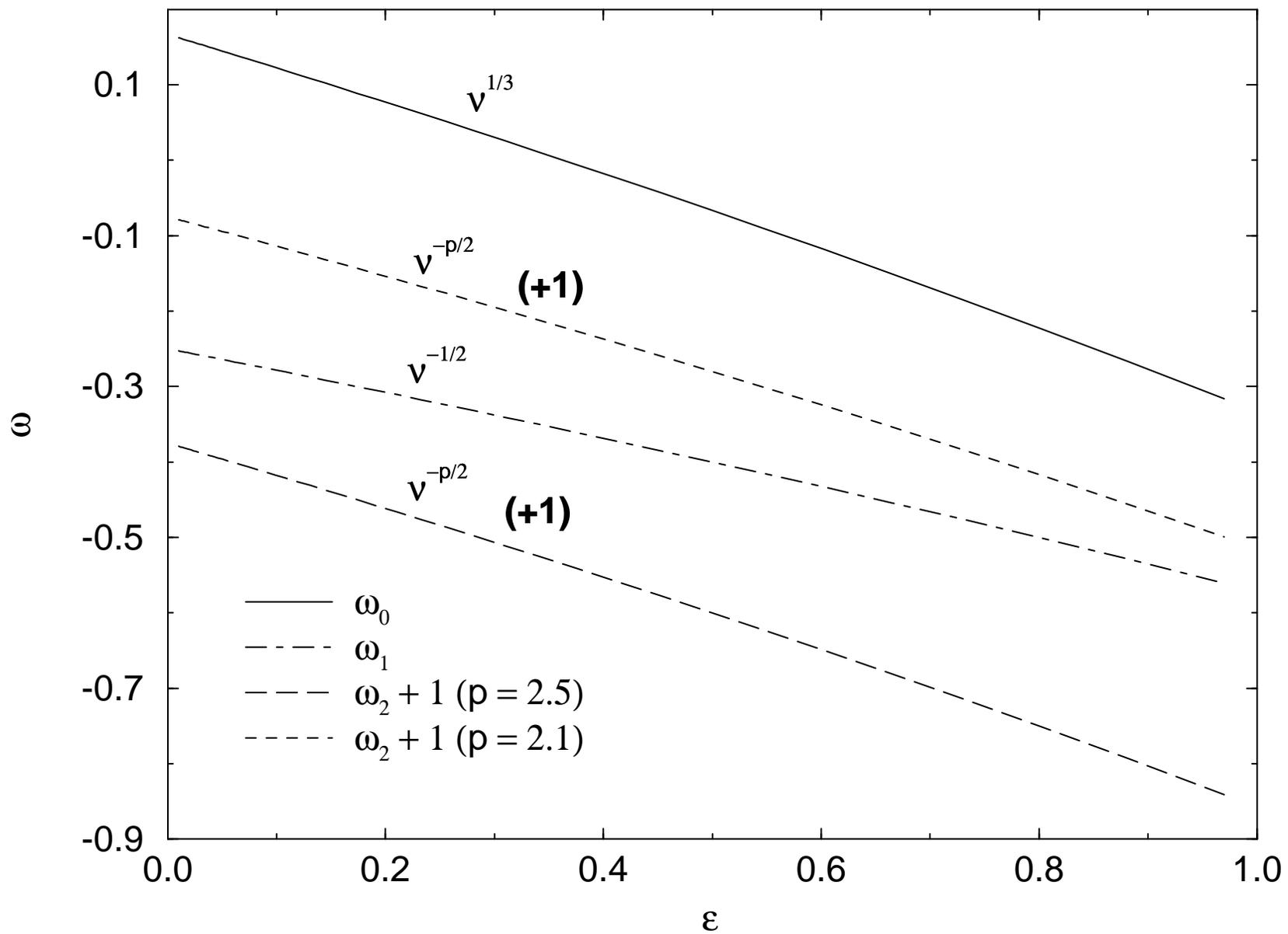
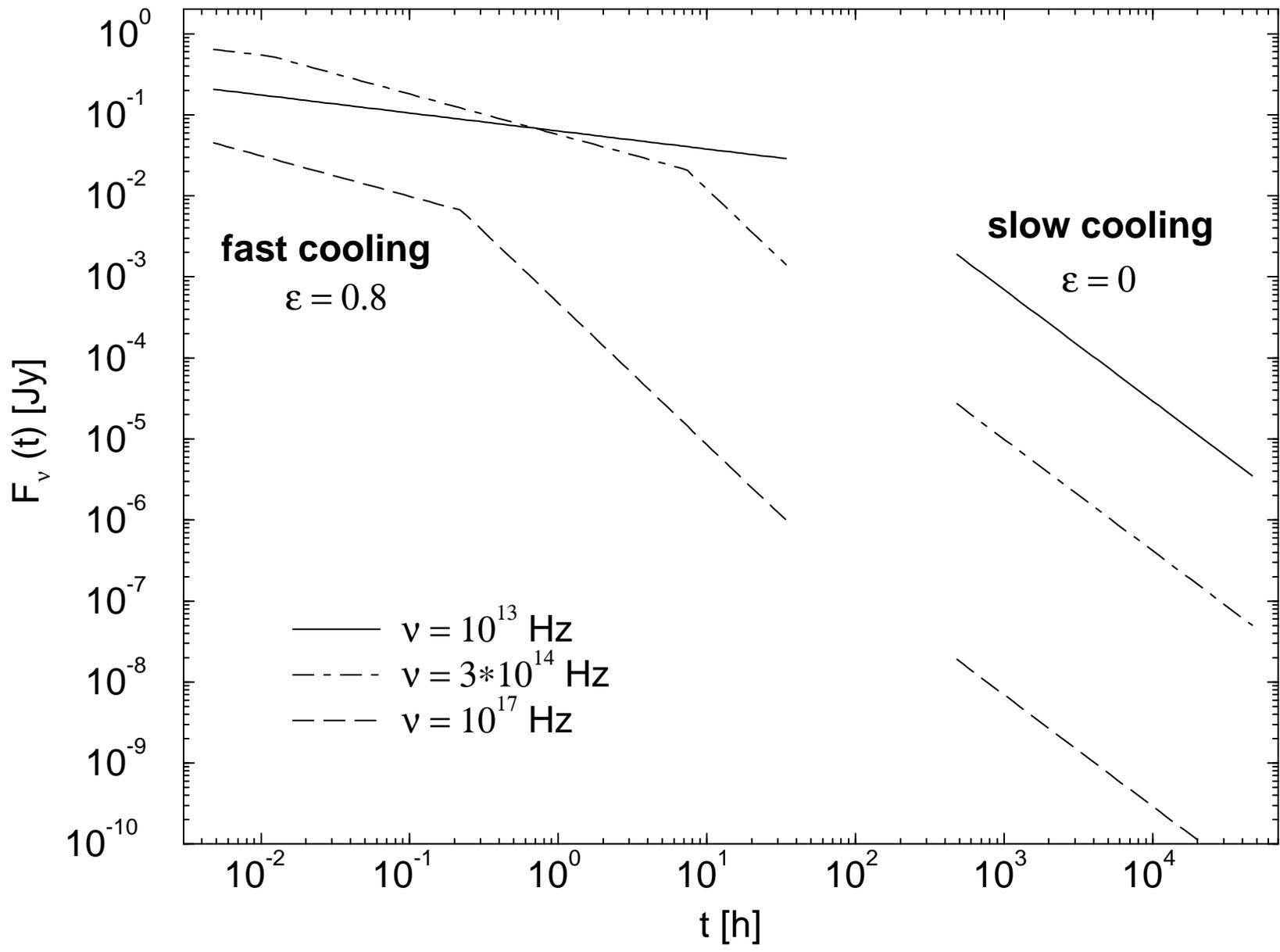
\end{document}